\documentclass[lettersize,journal]{IEEEtran}
\usepackage{amsfonts}
\usepackage{newtxtext}
\usepackage{amsmath}
\usepackage{algorithmic}
\usepackage{algorithm}
\usepackage{multirow}
\usepackage{booktabs}
\usepackage{array}
\usepackage[caption=false]{subfig}
\usepackage{textcomp}
\usepackage{adjustbox}
\usepackage{stfloats}
\usepackage{url}
\usepackage{verbatim}
\usepackage{graphicx}
\usepackage{caption}
\usepackage[colorlinks=true, linkcolor=blue, citecolor=blue, urlcolor=blue]{hyperref}
\usepackage{cite}
\usepackage{multirow}
\usepackage{colortbl}
\usepackage{graphicx}
\usepackage{threeparttable}
\usepackage{orcidlink}

\usepackage[table,xcdraw]{xcolor}
\usepackage[normalem]{ulem}
\usepackage{titlesec}
\titlespacing{\subsection}{0pt}{*0.5}{*0.5}

\useunder{\uline}{\ul}{}
\begin{document}

\title{MDFI: A Multi-Domain Features Integration for Compressed Video Quality Enhancement}

\author{
Sang NguyenQuang, 
Hieu Bui Minh, 
Dang BuiDinh, 
and Xiem HoangVan
\thanks{Sang NguyenQuang is with the Department of Computer Science, National Yang Ming Chiao Tung University, Hsinchu 30010, Taiwan.}
\thanks{Xiem HoangVan, Hieu Bui Minh and Dang BuiDinh are with the Faculty of Electronics and Telecommunications, VNU-University of Engineering and Technology, Vietnam National University, Hanoi 100000, Vietnam.}
\thanks{Corresponding author: Xiem HoangVan (e-mail: xiemhoang@vnu.edu.vn).}
}


\maketitle

\begin{abstract}
The latest video coding standard, H.266/VVC, has demonstrated significant improvements in compression efficiency compared to H.265/HEVC. Despite its advanced coding techniques, H.266/VVC still faces challenges in meeting the increasing demand for higher perceptual quality and enhanced compression performance. To address these limitations, we propose MDFI (Multi-Domain Features Integration), a compressed video quality enhancement approach that features a novel Frame-Prediction Feature Transform (FPFT) module to process prediction information. Moreover, MDFI integrates a multi-domain feature fusion strategy that effectively combines spatiotemporal characteristics, cross-frequency representations, and compressed-domain prediction information to enhance decoded video quality. Additionally, we introduce a comprehensive dataset that encompasses uncompressed video sequences, corresponding reconstructed versions at multiple QP levels, and predicted frames generated from H.266/VVC compressed bitstreams, providing essential resources for developing and benchmarking video enhancement approaches. Extensive experiments demonstrate that our MDFI approach achieves superior performance to state-of-the-art methods in both objective metrics and visual quality, effectively mitigating video compression artifacts.
The code is available at: \href{https://github.com/dangdinh17/MDFI.git}{https://github.com/dangdinh17/MDFI.git}
\end{abstract}


%
\IEEEpeerreviewmaketitle

\section{Introduction}
 
\IEEEPARstart{T}{he} rapid growth of UHD, 360°, and XR video applications has intensified the demand for efficient compression. The latest video coding standard, H.266/Versatile Video Coding (H.266/VVC)~\cite{overview_VVC}, represents a significant advancement over its predecessor, H.265/High Efficiency Video Coding (H.265/HEVC)~\cite{overview_HEVC}, achieving approximately 50\% bitrate reduction while maintaining comparable perceptual quality. This improvement is realized through advanced coding techniques, including larger and more flexible block partitioning structures, enhanced intra- and inter-prediction modes, and improved transform and entropy coding methods. However, despite its advanced coding tools, H.266/VVC still produces noticeable artifacts, especially blocking and blurring at low bitrates, which traditional filters cannot fully remove. These compression artifacts not only degrade visual quality but also negatively impact user experience and the performance of downstream computer vision tasks. {These challenges are critical in applications such as video post-production, streaming, and content restoration, where high visual quality must be maintained under aggressive compression. Effective artifact removal is therefore essential to enhance visual quality and improve the reliability of downstream processing tasks.}

In recent years, deep learning-based approaches have been widely adopted for video quality enhancement. Early methods for mitigating quality degradation in compressed video primarily adapted single-image enhancement techniques~\cite{ARCNN, qecnn, du2020blind}. These methods consider spatial correlation in only a single frame, ignoring effective information from adjacent neighbors; therefore, their performance is inherently limited. Later approaches~\cite{mfqr, mfqe, mfqev2, stdf, rfda, tgaf, tvqe,ovqe, stff, ovqe2024atc, cpga, cvqe2025atc} leveraged inter-frame correlations by exploiting spatiotemporal information throughout the sequence, achieving improved reconstruction quality and more consistent frame enhancement. Recently, researchers have explored incorporating compression-related information by leveraging auxiliary cues from other domains to further improve enhancement performance. 
In compressed video quality enhancement, OVQE~\cite{ovqe} exploited omni-frequency information through grid-based propagation across the entire sequence, while CPGA~\cite{cpga} introduced a coding priors-guided aggregation network utilizing motion vectors, predicted frames, and residual frames. This strategy is efficient since these priors are inherently embedded in the video bitstream.

\begin{figure}
    \centering
    \includegraphics[width=0.95\linewidth]{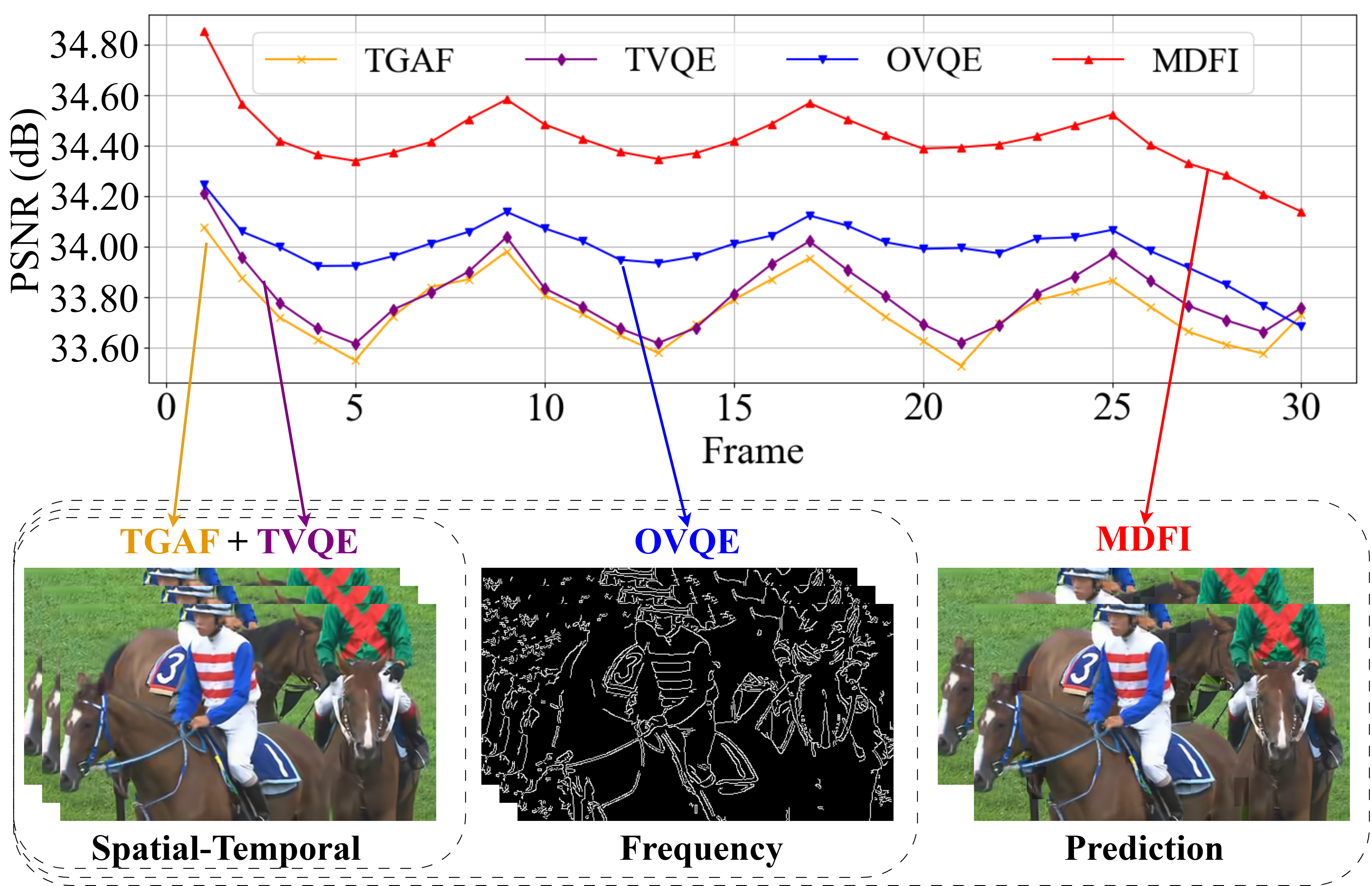}
    \vspace{0.2cm}
    \caption{PSNR comparison and exploited information in TGAF~\cite{tgaf}, TVQE~\cite{tvqe}, OVQE~\cite{ovqe} and our MDFI.}
    \label{fig:info}
    \vspace{-0.5cm}
\end{figure}

Despite significant progress in compressed video quality enhancement, existing methods still suffer from several limitations. First, they fail to fully exploit the rich information available in the compressed bitstream, particularly prediction information. Second, there is a lack of specialized approaches designed for H.266/VVC, as most existing methods focus on earlier coding standards. Third, the absence of comprehensive datasets that provide access to internal encoding information hinders the development of learning-based methods capable of effectively leveraging compressed-domain features.

To overcome these limitations, we propose the Multi-Domain Feature Integration (MDFI) framework, a novel approach that jointly leverages prediction, spatiotemporal, and frequency domain information for compressed video quality enhancement. In contrast to transform-domain iterative reconstruction approaches~\cite{reviewer2}, MDFI performs prediction-guided feature learning in the pixel domain without requiring decoder modification. The key innovation of our approach is the Frame-Prediction Feature Transform (FPFT) module, which leverages side information extracted from the H.266/VVC bitstream to enable prediction-guided feature alignment and temporal refinement. Built upon a multi-recursive propagation architecture, MDFI establishes long-range dependencies across frames and domains, achieving superior reconstruction quality and outperforming state-of-the-art methods~\cite{tvqe,tgaf, ovqe}, as shown in Fig.~\ref{fig:info}.
The main contributions of this work are summarized as follows:
\begin{itemize}
    \item A novel Multi-Domain Feature Integration (MDFI) framework that jointly leverages prediction signals, spatiotemporal characteristics, and frequency information via a multi-recursive propagation architecture, achieving superior performance and temporal consistency in compressed video quality enhancement.
    \item A novel Frame-Prediction Feature Transform (FPFT) module that processes prediction signals extracted from the bitstream and combines them with features from current frames through adaptive spatial feature fusion.
    \item A comprehensive dataset comprising raw sequences, decoded videos at various quality levels, and prediction frames extracted from H.266/VVC bitstreams. It provides essential multi-domain training data with access to internal coding information, facilitating the development and evaluation of advanced learning-based methods for compressed video quality enhancement research.
\end{itemize}


\section{Related works} \label{sec:related_work}

{Mitigating compression artifacts is vital to video quality and user experience. Beyond perceptual restoration, such enhancement techniques also benefit downstream multimedia tasks that handle complex visual distortions, including robust screenshot watermarking~\cite{TCSVT_Flexible, TMM_Screen, TCSVT_WaveRecovery} and forgery detection~\cite{ACM_PRest}. Existing methods for compressed video quality enhancement generally can be broadly categorized into three main groups: spatial, spatiotemporal, and multi-domain methods.}

\subsection{Spatial Methods}

Early approaches for mitigating quality degradation in compressed videos were directly adapted from single-image enhancement models. These methods operate on a frame-by-frame basis and treat each video frame independently, without considering temporal correlations. Typical tasks addressed by these approaches include image denoising~\cite{du2020blind}, general compression artifact removal~\cite{ARCNN}, and deblocking~\cite{li2017efficient}. The core strategy of these models is to exploit spatial correlations within an individual frame by learning mappings from degraded images to high-quality reconstructions using convolutional neural networks.

To further improve enhancement performance, several advanced single-frame architectures began incorporating information from multiple representation domains. For example, models such as D3~\cite{d3} and DualBDNet~\cite{du2020blind} leverage both pixel-domain features and frequency-domain features extracted from the DCT coefficients, enabling the networks to better characterize compression artifacts that are inherently generated in the transform domain. In addition, specialized quality enhancement models were designed to handle different frame types separately. For instance, QE-CNN~\cite{qecnn} adopts distinct sub-networks for I-frames and P/B-frames, reflecting the fact that different frame types exhibit different artifact characteristics.

Despite these architectural improvements, all single-frame spatial methods share a fundamental limitation: they inherently ignore the valuable spatiotemporal information that exists between adjacent frames. As a result, these methods are unable to exploit inter-frame redundancy or motion continuity, which are critical for restoring temporally coherent video sequences. This limitation becomes particularly severe in scenarios involving complex motion, rapid scene changes, or long-term temporal dependencies, where purely spatial models often produce temporal flickering and inconsistent visual quality across frames.

\subsection{Spatiotemporal Methods}

Spatiotemporal methods improve the performance of deep learning models by using multiple adjacent frames as input, rather than just a single frame, to extract correlations between spatial and temporal information. The first such method proposed was MFQE~\cite{mfqe}, which used a Multi-Frame Convolutional Neural Network (CNN). Shortly after, its successor, MFQEv2.0~\cite{mfqev2}, introduced a Bidirectional Long Short-Term Memory network for its Peak Quality Frames detector and became a benchmark for future work. A significant innovation in compressed video quality enhancement came with the Deformable Convolutional Network (DCN)~\cite{dcn}. Its deformable sampling architecture allowed models to capture complex motion across multiple frames more effectively than standard CNNs. This advancement led to better performance in subsequent models, such as TGAF~\cite{tgaf} and TVQE~\cite{tvqe}.

Another approach involves recursive-based models, which employ a recursive fusion module to propagate temporal correlations between frames. The state-of-the-art model RFDA~\cite{rfda} utilizes a Recursive Fusion module based on deformable convolution. It recursively combines previously compensated features with current ones, achieving large performance gains. Although these methods are fast, their performance is limited because they do not exploit the spatiotemporal information of the entire video. To address this limitation, recent models like OVQE~\cite{ovqe} and Wang et al.~\cite{stff} use a bidirectional, multi-stage recursive propagation scheme. This allows the current frame's features to utilize all refined features —past, present, and future— during each propagation pass, enabling these methods to achieve state-of-the-art results in compressed video quality enhancement.




\subsection{Multi-Domain Methods}

In recent years, multi-domain methods have attracted increasing attention in video enhancement research. Although spatiotemporal deep learning models have achieved remarkable performance improvements, many of them operate solely on decoded video frames and neglect the rich coding information embedded in compressed video bitstreams. This omission can limit further performance gains, as compressed bitstreams naturally contain valuable temporal and spatial priors introduced during the video encoding process.

Such coding priors include motion vectors, prediction modes, residual signals, and reference frame information, all of which provide explicit cues about inter-frame dependencies and compression distortions. Recently, several pioneering studies~\cite{cdvsr, cavsr} have demonstrated the effectiveness of incorporating codec-related information into video super-resolution frameworks, showing that coding priors can significantly enhance reconstruction quality when properly exploited. Inspired by these advances, CPGA~\cite{cpga} proposed a coding priors-guided aggregation network for the compressed video quality enhancement task. CPGA explicitly extracts and integrates three types of coding priors—motion vectors, predicted frames, and residual frames—to guide the enhancement process. By leveraging both pixel-domain information and codec-domain cues, this approach effectively bridges the gap between traditional signal processing and data-driven learning, achieving state-of-the-art performance. These multi-domain strategies highlight a promising research direction for further improving compressed video quality enhancement by jointly exploiting spatial, temporal, and coding-domain information.

\begin{figure*}[t]
    \centering
    \includegraphics[width=0.9\linewidth]{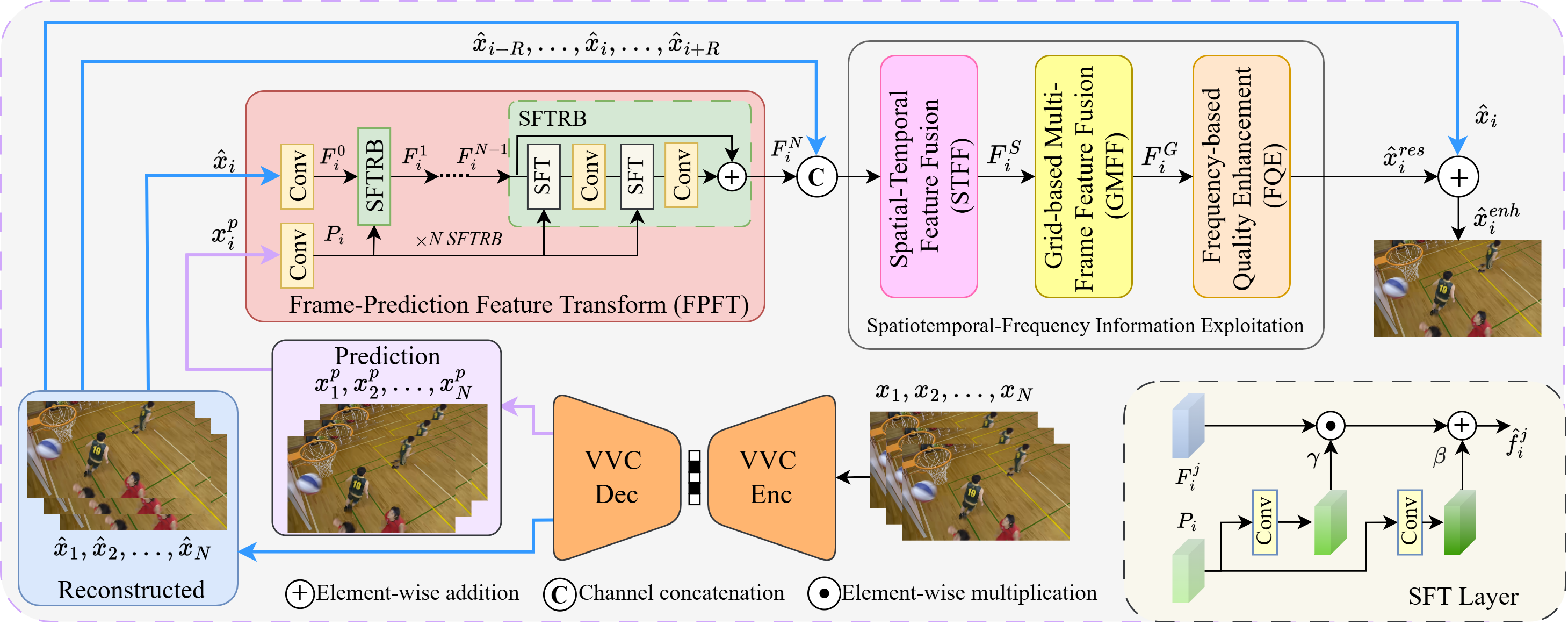}
    \centering
    \caption{{Overview of the MDFI architecture with the proposed Frame-Prediction Feature Transform. The full framework is illustrated in the gray background, while the SFT submodule is highlighted in the bottom-right corner.}}
    \label{fig:architecture}
    \vspace{-0.5cm}
\end{figure*}
\begin{figure}[!t]
    \centering
    \setlength{\tabcolsep}{1pt}
    \renewcommand{\arraystretch}{1}
    \begin{tabular}{ccc}

        \includegraphics[width=0.33\linewidth]{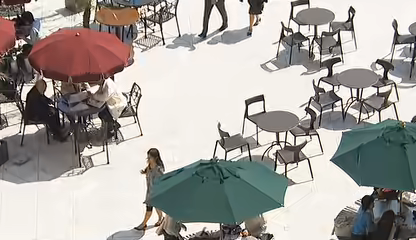} &
        \includegraphics[width=0.33\linewidth]{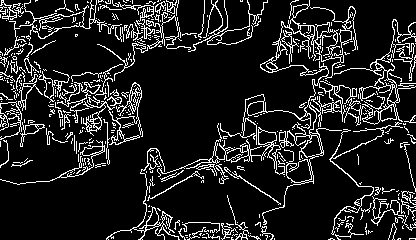} &
        \includegraphics[width=0.33\linewidth]{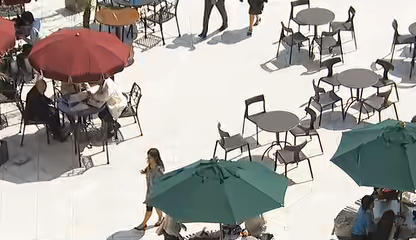} \\
        
        \includegraphics[width=0.33\linewidth]{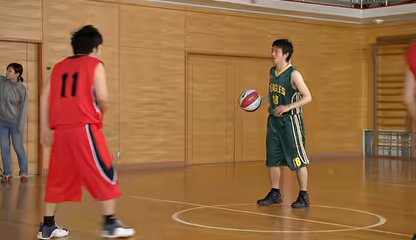} &
        \includegraphics[width=0.33\linewidth]{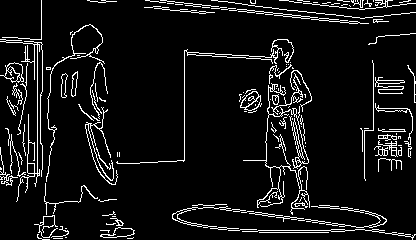} &
        \includegraphics[width=0.33\linewidth]{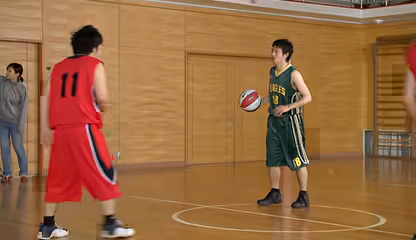} \\

    Input Frame &  Frequency &  Prediction  \\

    \end{tabular}
    \caption{Visualization of decoded information used in MDFI.}
    \label{fig:input_elements}
    \vspace{-0.5cm}
\end{figure}

\section{Proposed Method: MDFI}
\label{sec:proposed_method}

\subsection{System Overview}
\label{ssec:ovqe_vvc}
As illustrated in Fig.~\ref{fig:architecture}, the proposed MDFI framework exploits prediction signals extracted from compressed bitstreams, together with spatiotemporal and frequency information, to perform quality enhancement for H.266/VVC compressed videos.
Consider a video sequence of $T$ frames $\left\{ x_{1},x_{2},...,x_{T} \right\}$, where $x_{t}$ denotes the $t$-th frame. After compression by the H.266/VVC encoder and subsequent decoding, we obtain the reconstructed sequence $\left \{\hat{x}_{1}, \hat{x}_{2}, ..., \hat{x}_{T} \right \}$, where $\hat{x}_t$ represents the decoded version of $x_t$, along with the corresponding prediction information $\left \{ x_{1}^{p},x_{2}^{p},...,x_{T}^{p} \right \}$. This procedure can be formally represented as:
\begin{equation}
\label{equ:enc_dec_vvc_modified}
\left\{
\begin{aligned}
&\hat{x}_{1}, \hat{x}_{2},\ldots, \hat{x}_{T} \\
&x_{1}^{p}, x_{2}^{p},\ldots, x_{T}^{p}
\end{aligned}
\right\}
=
f_{dec}^{VVC} \left( f_{enc}^{VVC}(x_{1}, x_{2},\ldots, x_{T}) \right)
\end{equation}
where $f_{enc}^{VVC}(\cdot)$ and $f_{dec}^{VVC}(\cdot)$ represent the encoder and decoder of the H.266/VVC standard, respectively. Compared to the original uncompressed video, the reconstructed version naturally suffers from visual quality degradation, including blocking artifacts, blurring effects, and loss of fine textural details, due to the lossy compression process.

To further enhance the quality of a decoded video frame $\hat{x}_i$, we propose the multi-domain feature integration and temporal refinement framework, MDFI, which takes the target frame along with its temporal neighboring frames ($\hat{x}_{i-R}, ..., \hat{x}_{i}, ..., \hat{x}_{i+R}$ with $R=3$), as well as the corresponding prediction frame $x^p_i$ extracted from the bitstream. At the beginning of the pipeline, the \textit{Frame-Prediction Feature Transform} (FPFT) module processes the input frame $\hat{x}_i$ and its prediction frame $x_i^p$ using a sequence of Spatial Feature Transform ResBlocks (SFTRBs) to generate a comprehensive feature representation $F_{i}^{N}$. Afterward, the \textit{Spatio-Temporal Feature Fusion} (STFF) module aggregates the decoded frame $\hat{x}_i$ and the prediction features \( F_{i}^{p} \) along with its surrounding temporal context (i.e., decoded frames) ($\hat{x}_{i-R},..., \hat{x}_{i+R}$) to construct spatio-temporally fused features \( F_{i}^{S} \). These features are then refined by the \textit{Grid-based Multi-Frame Feature Fusion} (GMFF) module, which captures multi-scale dependencies, yielding the globally enriched representation \( F_{i}^{G} \). 
Subsequently, the \textit{Frequency-based Quality Enhancement} (FQE) module estimates the residual component \( x_{i}^{res} \) representing the discrepancy between the decoded frame and the high-quality frame. Finally, the enhanced frame \( x_{i}^{enh} \) is reconstructed by adding the predicted residual \( x_{i}^{res} \) to the decoded frame $\hat{x}_i$, thereby completing the enhancement process.
The overall process of our MDFI model can be summarized as:
\begin{align}
F_i^{N} &= FPFT(\hat{x}_i, x_i^{p}), \label{eq:fpft} \\
F_i^{S} &= STFF(\mathcal{C}(F_i^{p}, \hat{x}_{i-R:i+R})), \label{eq:stff} \\
F_i^{G} &= GMFF(F_i^{S}), \label{eq:gmff} \\
x_i^{res} &= FQE(F_i^{G}), \label{eq:qenet} \\
x_i^{enh} &= \hat{x}_i + x_i^{res}, \label{eq:enhanced}
\end{align}
where $\mathcal{C}(\cdot)$ denotes the concatenation operation, and $\hat{x}_{i-R:i+R}$ denotes the $2R+1$ adjacent decoded frames from time indices $i-R$ to $i+R$.
\subsection{Prediction Information Exploitation}
Prediction information in video compression serves as an estimation of the current block, derived from previously reconstructed samples via intra or inter prediction, and is recreated identically at both the encoder and decoder using control data (modes, motion vectors) parsed from the bitstream. At the decoder, the coding mode of each block allows reconstruction of the predicted frame, \( x_i^p \), which contains motion and texture cues that guided the original compression process. Fig.~\ref{fig:architecture} illustrates a sample video frame alongside its corresponding predictive information, represented as a predicted frame.

To fully exploit this valuable prediction information, we propose the FPFT module, which jointly processes the target decoded frame  $\hat{x}_i$ that requires quality enhancement and the corresponding prediction frame $x_i^p$ to generate its representation in the feature space \( F_i^p \), which serves as input for subsequent enhancement stages. By leveraging the prediction information, this module obtains reliable structural and motion priors that guide the network in restoring missing high-frequency details and achieving accurate temporal feature alignment across frames. 

As shown in Fig.~\ref{fig:architecture}, the FPFT module processes the decoded frame $\hat{x}_i$ and prediction frame $x_i^p$ through initial convolutional layers to extract feature maps, $F_i^0$ and $P_i$, respectively. The feature map $F_i^0$ then passes through $N$ sequential SFTRBs, each containing two Spatial Feature Transform (SFT) layers~\cite{sft} with residual connections, producing intermediate features \{$F_i^1,...,F_i^{N-1}$\} and output feature $F_i^N$. Within each SFT layer (detailed in the lower part of Fig.~\ref{fig:architecture}), spatially adaptive affine transformations modulate the input feature map based on $P_i$. Specifically, the modulation parameters $\gamma$ (scaling) and $\beta$ (shifting) are generated through convolutional layers:
\begin{equation}
\gamma = \text{Conv}_\gamma(P_i), \quad \beta = \text{Conv}_\beta(P_i).
\end{equation}
The transformed features $\hat{f}_i^j$ are then computed as:
\begin{equation}
\hat{f}_i^j =\gamma \odot F_i^j + \beta,
\end{equation}
where \( \odot \) denotes element-wise multiplication. This adaptive transformation enables prediction-guided feature refinement, thus improving alignment with decoded content and achieving targeted artifact suppression.



By stacking multiple such SFTRBs with SFT layers, the FPFT module progressively refines the joint representation $F_i^N = FPFT(\hat{x}_i, x_i^p)$, which is subsequently fed into the spatiotemporal and frequency-aware stages of the proposed MDFI framework. The use of SFT layers in this pipeline enhances the expressiveness of the feature extraction process while improving the model’s ability to preserve structural details and mitigate prediction-related distortions.

\subsection{Spatio–Temporal–Frequency Exploitation}
To further exploit spatiotemporal and frequency features, we adopt the feature extraction strategy from OVQE~\cite{ovqe}, which comprises three key components: STFF, GMFF, and FQE, as shown in Fig.~\ref{fig:architecture}. These modules serve as concrete implementations of the transformations defined in Eqs.~(\ref{eq:fpft})--(\ref{eq:enhanced}), where each component corresponds to a specific stage in the overall pipeline.
\begin{itemize}

\item \textbf{STFF:} Unlike the original STFF module in OVQE, which only processes the current frame $\hat{x}_i$ and its neighboring frames ($\hat{x}_{i-R}, \ldots, \hat{x}_{i+R}$), we additionally incorporate refined prediction features $F_i^p$ obtained from Eq.~(\ref{eq:fpft}).
Specifically, STFF takes the concatenated input $\mathcal{C}(F_i^{p}, \hat{x}_{i-R:i+R})$ and produces the spatiotemporal feature representation $F_i^{S}$ as defined in Eq.~(\ref{eq:stff}). The module adopts a U-Net-based architecture for multi-scale representation, where features are processed and then aggregated via Selective Kernel Feature Fusion (SKFF)~\cite{skff} modules to combine complementary information. The fused features are further refined by a DCN module, which samples from flexible spatial locations to capture richer spatiotemporal dependencies.

\item \textbf{GMFF:} 
This scheme is inspired by BasicVSR++~\cite{basicvsr++} and takes the spatiotemporal features $F_i^{S}$ from Eq.~(\ref{eq:stff}) as input to generate enhanced global features $F_i^{G}$, {corresponding to Eq.~(\ref{eq:gmff}). GMFF combines grid-based enhancement with bidirectional propagation.} Unlike BasicVSR++, GMFF integrates features from previous propagation stages, allowing each frame to utilize contextual information from past, present, and future frames. GMFF has four-stage bidirectional propagation (two forward, two backward). Each propagation stage sequentially applies STFF for spatiotemporal alignment and OFAE (Omni-Frequency Adaptive Enhancement) for omni-frequency enhancement, where OFAE effectively extracts and retains multi-scale frequency features throughout the sequence, ensuring robust temporal consistency.

\item \textbf{FQE:} 
This module {takes the propagated features $F_i^{G}$ from Eq.~(\ref{eq:gmff}) and estimates the residual component $x_i^{res}$ as defined in Eq.~(\ref{eq:qenet}). 
It is implemented as a lightweight quality enhancement network} with five OFAE blocks and two convolutional layers. Owing to the critical role of frequency information in video frames, OFAE modules enable FQE to reconstruct frequency components more accurately, thereby achieving perceptually consistent and high-fidelity residual enhancement. Finally, the enhanced frame $x_i^{enh}$ is reconstructed by adding the predicted residual $x_i^{res}$ to the decoded frame $\hat{x}_i$, as in Eq.~(\ref{eq:enhanced}).

\end{itemize}

\subsection{H.266/VVC-Compressed Dataset Construction}
\label{ssec:dataset}

In video codecs such as H.266/VVC, all compression-related information, including motion vectors, prediction modes, residual data, and reconstructed frames, is embedded in the bitstream. Therefore, we introduce a comprehensive dataset, MDFI\footnote{Our proposed MDFI dataset can be found at \href{https://www.kaggle.com/datasets/dangdinh123/mdfi-dataset}{https://www.kaggle.com/datasets/dangdinh123/mdfi-dataset}.}, which contains raw and reconstructed sequences along with the corresponding prediction information extracted from the compressed bitstream. Following MFQEv2~\cite{mfqev2}, we collect 126 video sequences from Xiph.org~\cite{xiph_video}, VQEG~\cite{vqeg_datasets}, and JCT-VC~\cite{jctvc}. These sequences are divided into a training set (108 sequences) and a testing set (18 sequences)\footnote{The original MFQEv2 dataset contains 160 videos (106 for training, 36 for validation, and 18 for testing). However, recent works commonly adopt a subset of 108 sequences for training and 18 videos for testing.}. Various types of scenes are included, encompassing a wide range of content categories such as human-centric scenes, natural environments, and urban landscapes, with resolutions ranging from SIF ($352\times240$), CIF ($352\times288$), NTSC ($720\times486$), 4CIF ($704\times576$), 240p ($416\times240$), 360p ($640\times360$), 480p ($832\times480$), 720p ($1280\times720$), and 1080p ($1920\times1080$) to WQXGA ($2560\times1600$). To enable research on quality enhancement methods for H.266/VVC compressed video, we encode all videos using the VVenC~\cite{VVenC} encoder under the Low Delay P (LDP) configuration at four QPs: 22, 27, 32, and 37. Comparisons between our MDFI dataset and the MFQEv2 dataset are presented in Table~\ref{tab:dataset}. {In addition, the proposed MDFI dataset is publicly available for academic and research purposes, with detailed usage instructions provided via the dataset link, facilitating reproducibility and future research.}




\section{Experiments} \label{sec:experiment}
\subsection{Implementation Details}

\textbf{Training details:} We train our models with the MDFI dataset, with the number of adjacent input frames set to 7 ($R=3$). During training, we randomly crop sub-frame clips of size $128 \times 128$ from raw videos, reconstructed videos, and predicted frames to generate training samples. Each sample consists of 15 consecutive frames. Data augmentation is performed using random flipping and rotation. We adopt the Adam optimizer~\cite{Adam} with $\beta_{1} = 0.9$, $\beta_{2} = 0.999$, $\epsilon = 10^{-6}$, and an initial learning rate of $1 \times 10^{-4}$, which linearly decays until convergence. In addition, we employ the Charbonnier loss~\cite{Charbonnier_Loss} as the final objective function. 



\begin{table}[!t]
\setlength{\tabcolsep}{2pt}

\centering
\caption{MFQEv2 dataset~\cite{mfqev2} versus our MDFI dataset.}
\label{tab:dataset}
\resizebox{\linewidth}{!}{%
%
\begin{tabular}{lcc}
\hline
     Dataset               & MFQEv2~\cite{mfqev2}             & MDFI (Ours)              \\ 
\hline
Total sequences     & 160                               & 126                      \\
Training sequences  & 106                               & 108                      \\
Validation sequences & 36                               & $\times$ \\
Testing sequences   & 18                                & 18                       \\
Predicted sequences & $\times$                          & \checkmark                       \\

{Diversity} & {Interview, Sports, Natural, Urban} & {Interview, Sports, Natural, Urban} \\
{Resolution range} & {From SIF to WQXGA} & {From SIF to WQXGA} \\

Compression codec   & H.265/HEVC    & H.266/VVC  \\
{Compression settings} & {Low Delay P} & {Low Delay P} \\
{Quantization Parameters (QPs)} & {22, 27, 32, 37, 42} & {22, 27, 32, 37} \\
Software            & HM 16.5                            & VVenC/VVdeC  \\
\hline
\end{tabular}%
}
\vspace{-0.7cm}

\end{table}

\textbf{Evaluation Methodologies:} Consistent with prior works~\cite{ovqe, stff, stdf}, we apply quality enhancement only on the Y-channel in the YUV 4:2:0 space, and report improvements in Peak Signal-to-Noise Ratio ($\Delta$PSNR) and Structural Similarity Index Measure~\cite{ssim} ($\Delta$SSIM). Rate–distortion performance is evaluated using BD-rate savings~\cite{bdrate}, with VVenC serving as the anchor. Model size and computational complexity are reported in terms of the number of parameters and FLOPs (floating-point operations) evaluated with respect to class E, respectively. {We compare our MDFI with state-of-the-art methods, including TGAF~\cite{tgaf}, TVQE~\cite{tvqe}, OVQE~\cite{ovqe}, and Wang et al~\cite{stff}.} All the baseline methods are finetuned on our proposed MDFI dataset to ensure a fair comparison. It can be noted that the best results in all tables are highlighted in \textbf{bold}. 

\subsection{Experimental Results}
\label{ssec:performance_evaluation}
\begin{figure}[t]
    \centering
    \begin{tabular}{c}
          \includegraphics[width=1\linewidth]{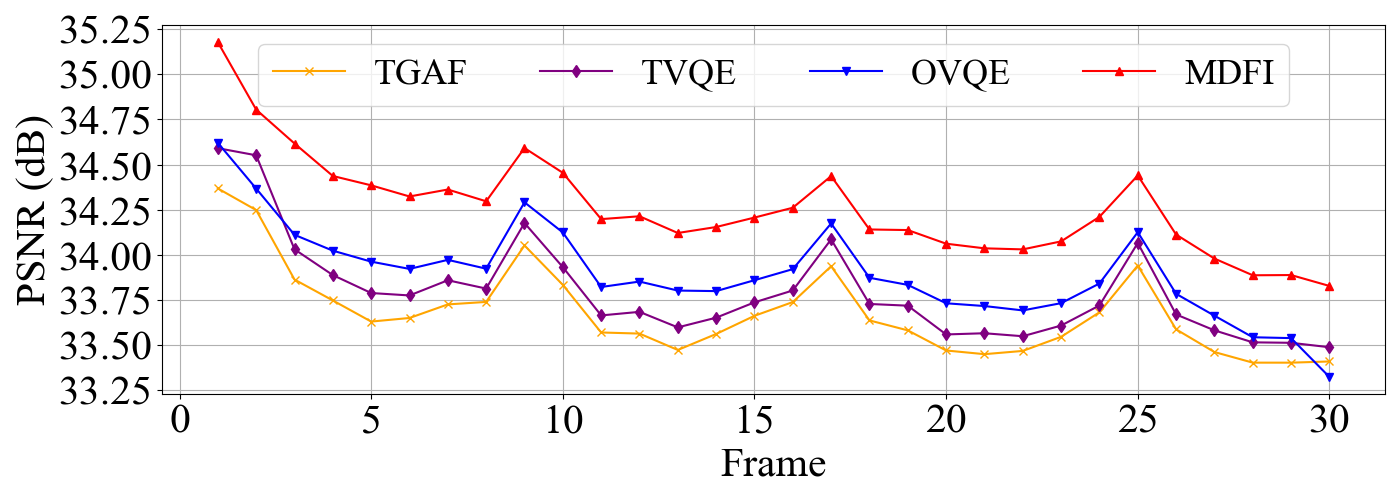}  \\
          \includegraphics[width=1\linewidth]{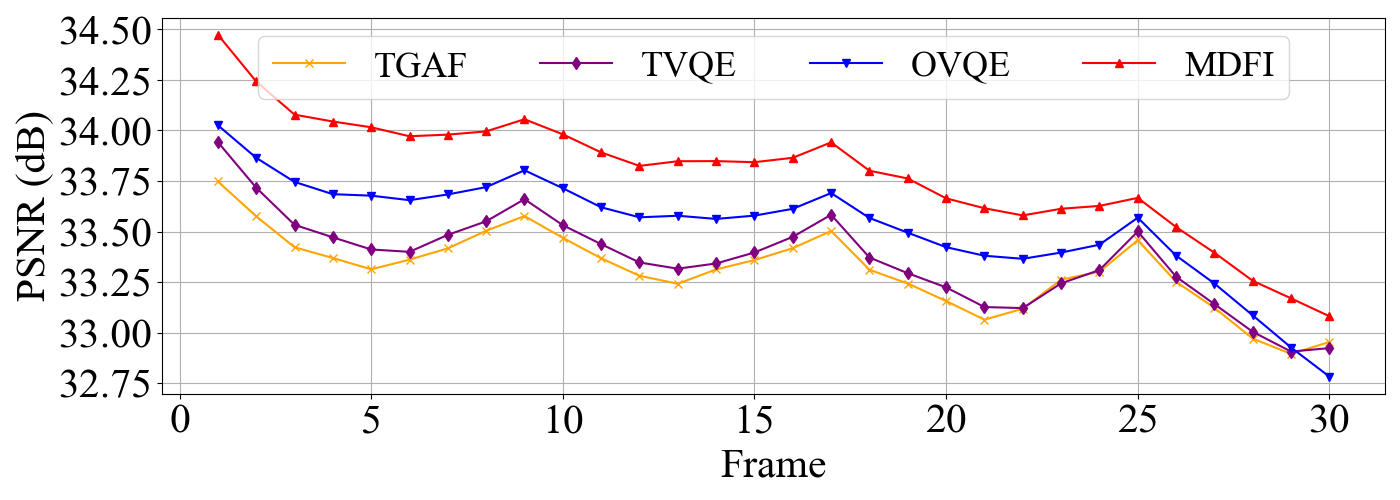}
    \end{tabular}
    \caption{PSNR comparison across frames on  BasketballDrill (top) and BQMall (bottom) sequences at QP = 37.}
    \label{fig:psnr_results}
    \vspace{-0.4cm}
\end{figure}

\textbf{Overall Quality Enhancement:} Table~\ref{tab:overal_results} compares the quality enhancement performance of our method with state-of-the-art methods in terms of $\Delta$PSNR and $\Delta$SSIM across five standard test classes (A to E). {As observed, the proposed MDFI, which exploits prediction–spatiotemporal–frequency information, consistently outperforms state-of-the-art methods using only spatiotemporal (TGAF~\cite{tgaf}, TVQE~\cite{tvqe}) or spatiotemporal-frequency cues (OVQE~\cite{ovqe}, Wang et al.~\cite{stff}). At QP = 37, MDFI achieves the best averages of 0.85 dB in $\Delta$PSNR and 1.68 in $\Delta$SSIM, surpassing OVQE and Wang et al. by ~15\% and ~13\%. Compared with transformer-based TVQE and iterative TGAF, MDFI improves PSNR by 0.25 dB and 0.23 dB, confirming superior artifact reduction and structural preservation under medium–high compression.} Furthermore, the frame-wise PSNR curves in Fig.~\ref{fig:psnr_results} show that MDFI consistently achieves higher enhancement gains across frames with reduced temporal fluctuation, resulting in more stable performance and improved temporal consistency.

\begin{table*}[!t]
\centering
\scriptsize
\caption{Overall comparison for $\Delta$PSNR (dB) and $\Delta$SSIM ($\times 10^{-2}$) over standard test sequences at four QPs}
\label{tab:overal_results}
%
\begin{threeparttable}
\begin{tabular}{ccccccccccccc} 
\hline
\multirow{2}{*}{\textbf{QP}}  & \multirow{2}{*}{Class} & \multirow{2}{*}{Sequence} 
& \multicolumn{2}{c}{TGAF~\cite{tgaf}} 
& \multicolumn{2}{c}{TVQE~\cite{tvqe}} 
& \multicolumn{2}{c}{OVQE~\cite{ovqe}} 
& \multicolumn{2}{c}{{Wang et al~\cite{stff}}} 
& \multicolumn{2}{c}{MDFI (Ours)}  \\
& & 
& PSNR & SSIM 
& PSNR & SSIM 
& PSNR & SSIM 
& {PSNR} & {SSIM} 
& PSNR & SSIM \\ 
\hline

\multirow{19}{*}{\textbf{37}} 
& \multirow{2}{*}{\begin{tabular}[c]{@{}c@{}}A \\ (2560$\times$1600)\end{tabular}} 
& Traffic 
& 0.67 & 0.94 
& 0.67 & 0.97 
& 0.87 & 1.21 
& {0.90} & {1.25} 
& \textbf{1.04} & \textbf{1.40} \\

& & PeopleOnStreet 
& 0.96 & 1.80 
& 0.90 & 1.72 
& 1.02 & 1.91 
& {1.06} & {2.01} 
& \textbf{1.21} & \textbf{2.22} \\ 
\cline{2-13}

& \multirow{5}{*}{\begin{tabular}[c]{@{}c@{}}B \\ (1920$\times$1080)\end{tabular}} 
& Kimono 
& 0.72 & 1.27 
& 0.70 & 1.17 
& 0.88 & 1.52 
& {0.87} & {1.52} 
& \textbf{1.07} & \textbf{1.81} \\

& & ParkScene 
& 0.59 & 1.47 
& 0.53 & 1.34 
& 0.73 & 1.85 
& {0.76} & {1.97} 
& \textbf{0.89} & \textbf{2.18} \\

& & Cactus 
& 0.51 & 1.00 
& 0.51 & 0.93 
& 0.61 & 1.22 
& {0.62} & {1.21} 
& \textbf{0.71} & \textbf{1.35} \\

& & BQTerrace 
& 0.33 & 0.57 
& 0.34 & 0.65 
& 0.46 & 0.81 
& {0.41} & {0.70} 
& \textbf{0.51} & \textbf{0.86} \\

& & BasketballDrive 
& 0.50 & 0.86 
& 0.48 & 0.81 
& 0.63 & 1.04 
& {0.64} & {1.06} 
& \textbf{0.76} & \textbf{1.23} \\ 
\cline{2-13}

& \multirow{4}{*}{\begin{tabular}[c]{@{}c@{}}C \\ (832$\times$480)\end{tabular}} 
& RaceHorses 
& 0.35 & 1.06 
& 0.31 & 0.94 
& 0.39 & 1.27 
& {0.39} & {1.27} 
& \textbf{0.44} & \textbf{1.50} \\

& & BQMall 
& 0.82 & 1.60 
& 0.75 & 1.52 
& 0.96 & 1.88 
& {1.00} & {1.94} 
& \textbf{1.14} & \textbf{2.15} \\

& & PartyScene 
& 0.47 & 1.60 
& 0.45 & 1.54 
& 0.57 & 2.09 
& {0.57} & {1.95} 
& \textbf{0.60} & \textbf{2.17} \\

& & BasketballDrill 
& 0.45 & 0.88 
& 0.48 & 0.98 
& 0.53 & 1.02 
& {0.52} & {0.99} 
& \textbf{0.63} & \textbf{1.22} \\ 
\cline{2-13}

& \multirow{4}{*}{\begin{tabular}[c]{@{}c@{}}D \\ (416$\times$240)\end{tabular}} 
& RaceHorses 
& 0.59 & 1.79 
& 0.52 & 1.58 
& 0.65 & 2.05 
& {0.68} & {2.10} 
& \textbf{0.75} & \textbf{2.42} \\

& & BQSquare 
& 0.73 & 1.26 
& 0.65 & 1.22 
& \textbf{0.94} & \textbf{1.65} 
& {0.89} & {1.43} 
& 0.89 & 1.57 \\

& & BlowingBubbles 
& 0.55 & 2.12 
& 0.54 & 2.02 
& 0.71 & 2.80 
& {0.71} & {2.72} 
& \textbf{0.79} & \textbf{3.06} \\

& & BasketballPass 
& 0.85 & 1.99 
& 0.79 & 1.84 
& 0.93 & 2.25 
& {0.98} & {2.26} 
& \textbf{1.10} & \textbf{2.65} \\ 
\cline{2-13}

& \multirow{3}{*}{\begin{tabular}[c]{@{}c@{}}E \\ (1280$\times$720)\end{tabular}} 
& FourPeople 
& 0.77 & 0.85 
& 0.81 & 0.90 
& 0.96 & 1.00 
& {0.98} & {1.03} 
& \textbf{1.06} & \textbf{1.10} \\

& & Johnny 
& 0.52 & 0.41 
& 0.66 & \textbf{0.59} 
& 0.68 & 0.52 
& {0.69} & {0.54} 
& \textbf{0.76} & 0.48 \\

& & KristenAndSara 
& 0.70 & 0.61 
& 0.78 & 0.67 
& 0.87 & 0.73 
& {0.86} & {0.73} 
& \textbf{0.98} & \textbf{0.80} \\ 
\cline{2-13}

& \multicolumn{2}{c}{Average} 
& 0.62 & 1.23 
& 0.60 & 1.19 
& 0.74 & 1.49 
& {0.75} & {1.48} 
& \textbf{0.85} & \textbf{1.68} \\ 
\hline

\textbf{32} & \multicolumn{2}{c}{Average} 
& 0.67 & 0.99 
& 0.57 & 0.84 
& 0.82 & 1.22 
& {0.80} & {1.20} 
& \textbf{0.93} & \textbf{1.40} \\ 
\hline

\textbf{27} & \multicolumn{2}{c}{Average} 
& 0.56 & 0.62 
& 0.51 & 0.54 
& 0.76 & 0.82 
& {0.82} & {0.85} 
& \textbf{0.98} & \textbf{1.01} \\ 
\hline

\textbf{22} & \multicolumn{2}{c}{Average} 
& 0.62 & 0.39 
& 0.39 & 0.28 
& 0.72 & 0.48 
& {0.72} & {0.47} 
& \textbf{0.88} & \textbf{0.57} \\
\hline
\end{tabular}

\end{threeparttable}
\end{table*} 

\begin{figure*}[!t]
    \captionsetup[subfloat]{labelformat=empty}
    \begin{center}
        \newcommand{\rowArg}{1.75cm}      
        \newcommand{\fullheight}{4cm}
        \newcommand{\fullwidth}{8cm}
        \newcommand{\patchwidth}{2.5cm}  
        \newcommand{\patchheight}{1.7cm}
        \setlength\tabcolsep{0.1cm}      
        
        \begin{tabular}[]{c c c c}
            
            \multirow{2}{*}[\rowArg]{
                \subfloat[Compressed Frame] 
                {\includegraphics[width = \fullwidth, height = \fullheight]
                    {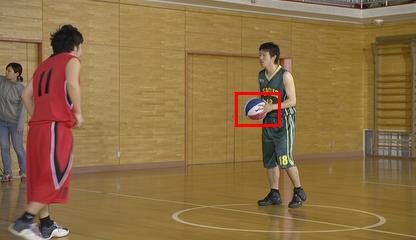}}} &
            
            \subfloat[H.266/VVC~\cite{overview_VVC}]
            {\includegraphics[width = \patchwidth, height = \patchheight]
                {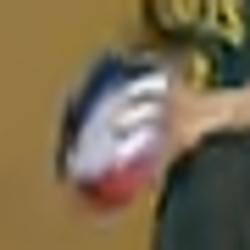}} &
            \subfloat[TGAF~\cite{tgaf}]
            {\includegraphics[width = \patchwidth, height = \patchheight]
                {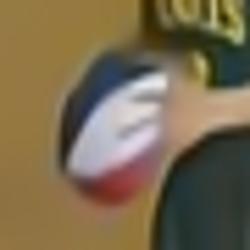}} &
            \subfloat[TVQE~\cite{tvqe}]
            {\includegraphics[width = \patchwidth, height = \patchheight]
                {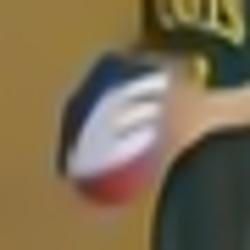}} \\[-0.2cm]&
            
            
            \subfloat[OVQE~\cite{ovqe}]
            {\includegraphics[width = \patchwidth, height = \patchheight]
                {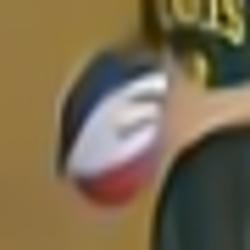}} &
            \subfloat[MDFI (Ours)] 
            {\includegraphics[width = \patchwidth, height = \patchheight]
                {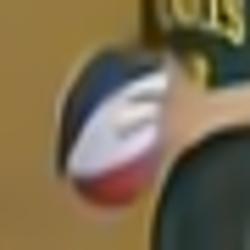}} &
            \subfloat[Raw]
            {\includegraphics[width = \patchwidth, height = \patchheight]
                {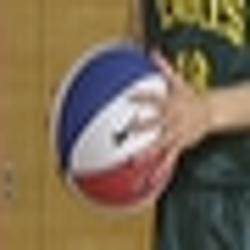}} \\

            \multirow{2}{*}[\rowArg]{
                \subfloat[Compressed Frame]
                {\includegraphics[width = \fullwidth, height = \fullheight]
                    {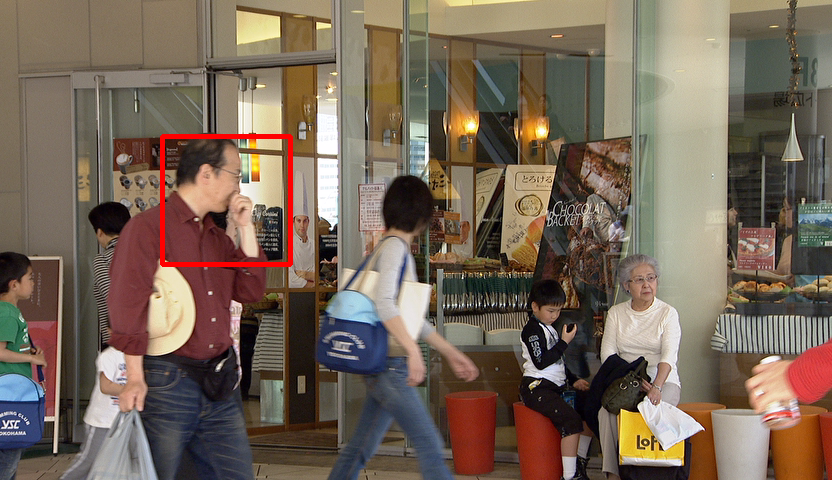}}} &
    
            \subfloat[H.266/VVC~\cite{overview_VVC}]
            {\includegraphics[width = \patchwidth, height = \patchheight]
                {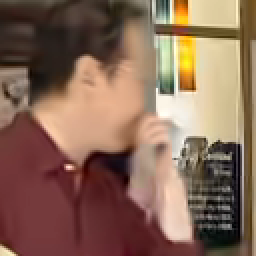}} &
            \subfloat[TGAF~\cite{tgaf}]
            {\includegraphics[width = \patchwidth, height = \patchheight]
                {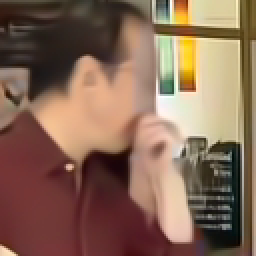}} &
            \subfloat[TVQE~\cite{tvqe}]
            {\includegraphics[width = \patchwidth, height = \patchheight]
                {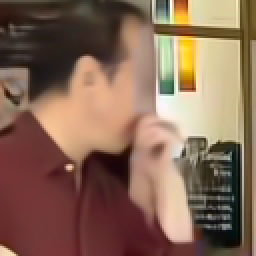}} \\[-0.2cm]&

            \subfloat[OVQE~\cite{ovqe}]
            {\includegraphics[width = \patchwidth, height = \patchheight]
                {figure/visual_comparison/BQMall_832x480_600_TVQE.png}} &
            \subfloat[MDFI (Ours)]
            {\includegraphics[width = \patchwidth, height = \patchheight]
                {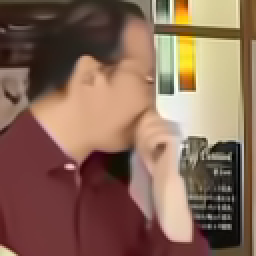}} &
            \subfloat[Raw]
            {\includegraphics[width = \patchwidth, height = \patchheight]
                {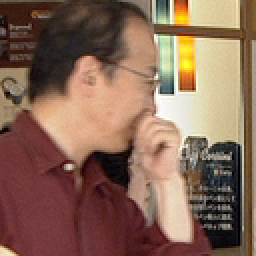}} \\ 

        \end{tabular}
    \end{center}
    \caption{Visual comparison of results between experimented methods. The sources of the video frames in the visualization results are: BasketballPass (416 × 240), BQMall (832 × 480).}
    \label{fig:visual_comparison}
    \vspace{-0.5cm}

\end{figure*}

\textbf{Subjective Quality Performance:} Fig.~\ref{fig:visual_comparison} visualizes the subjective quality comparison among H.266/VVC, TGAF~\cite{tgaf}, TVQE~\cite{tvqe}, OVQE~\cite{ovqe}, and the proposed MDFI at QP = 37 on representative sequences from different classes, including \textit{BasketballPass} and \textit{BQMall}. Overall, the qualitative results indicate that MDFI achieves superior visual quality by better preserving fine details, structural boundaries, and texture consistency across different content types. 
For the \textit{BasketballPass} sequence (top row), competing methods tend to over-smooth the hand–ball boundaries, while MDFI preserves sharper finger contours that are closer to the raw reference. In the \textit{BQMall} sequence (bottom row), all competing methods over-smooth frame of the glasses into the skin of the man's face, whereas MDFI realistically preserves these details. These qualitative results further demonstrate that MDFI achieves superior visual quality and more effective artifact suppression across various challenging scenarios compared to existing methods.

    
    
    
    
    
    

\begin{table}[!t]
\centering
\scriptsize
\caption{BD-rate $(\%)$ with H.266/VVC as the anchor.}
\label{tab:BDrate}
%

\resizebox{\linewidth}{!}{%

\begin{tabular}{l|lccccc} 
\hline
\multicolumn{2}{c}{\textbf{Sequence}} & \textbf{TGAF~\cite{tgaf}} & \textbf{TVQE~\cite{tvqe}} & \textbf{OVQE~\cite{ovqe}} & \textbf{{Wang et al~\cite{stff}}} & \textbf{MDFI}  \\ 
\hline
\multirow{2}{*}{A} & Traffic          & -18.31 & -18.59 & -25.15 & {-25.50} & \textbf{-29.23} \\
                   & PeopleOnStreet   & -13.39 & -12.48 & -15.00 & {-15.82} & \textbf{-18.91} \\ 
\hline
\multirow{5}{*}{B} & Kimono           & -17.31 & -15.06 & -20.53 & {-19.86} & \textbf{-24.83} \\
                   & ParkScene        & -20.60 & -17.34 & -26.77 & {-27.16} & \textbf{-30.80} \\
                   & Cactus           & -15.12 & -14.08 & -19.09 & {-19.11} & \textbf{-22.45} \\
                   & BQTerrace        & -12.71 & -12.39 & -18.64 & {-18.81} & \textbf{-23.46} \\
                   & BasketballDrive  & -13.74 & -11.40 & -17.09 & {-17.12} & \textbf{-21.69} \\ 
\hline
\multirow{4}{*}{C} & RaceHorses       & -9.73  & -6.46  & -10.35 & {-11.01} & \textbf{-13.94} \\
                   & BQMall           & -18.84 & -14.02 & -22.43 & {-23.42} & \textbf{-27.22} \\
                   & PartyScene       & -19.04 & -14.62 & -22.98 & {-23.35} & \textbf{-25.46} \\
                   & BasketballDrill  & -8.83  & -9.74  & -13.17 & {-13.84} & \textbf{-16.21} \\ 
\hline
\multirow{4}{*}{D} & RaceHorses       & -13.45 & -10.57 & -16.39 & {-17.27} & \textbf{-20.73} \\
                   & BQSquare         & -29.81 & -21.98 & -34.96 & {-35.31} & \textbf{-37.44} \\
                   & BlowingBubbles   & -20.86 & -17.40 & -26.87 & {-26.91} & \textbf{-29.72} \\
                   & BasketballPass   & -17.70 & -14.32 & -19.78 & {-20.64} & \textbf{-23.37} \\ 
\hline
\multirow{3}{*}{E} & FourPeople       & -18.16 & -19.43 & -23.18 & {-23.47} & \textbf{-26.45} \\
                   & Johnny           & -16.31 & -18.81 & -21.90 & {-21.79} & \textbf{-24.90} \\
                   & KristenAndSara   & -18.93 & -20.41 & -24.21 & {-23.97} & \textbf{-27.12} \\ 
\hline
\multicolumn{2}{c}{Average} & -16.82 & -14.95 & -21.03 & {-21.35} & \textbf{-24.66} \\ 
\hline    
\end{tabular}
}
\vspace{-0.4cm}
\end{table}

\begin{figure}[!t]
    \centering
    \includegraphics[width=1\linewidth]{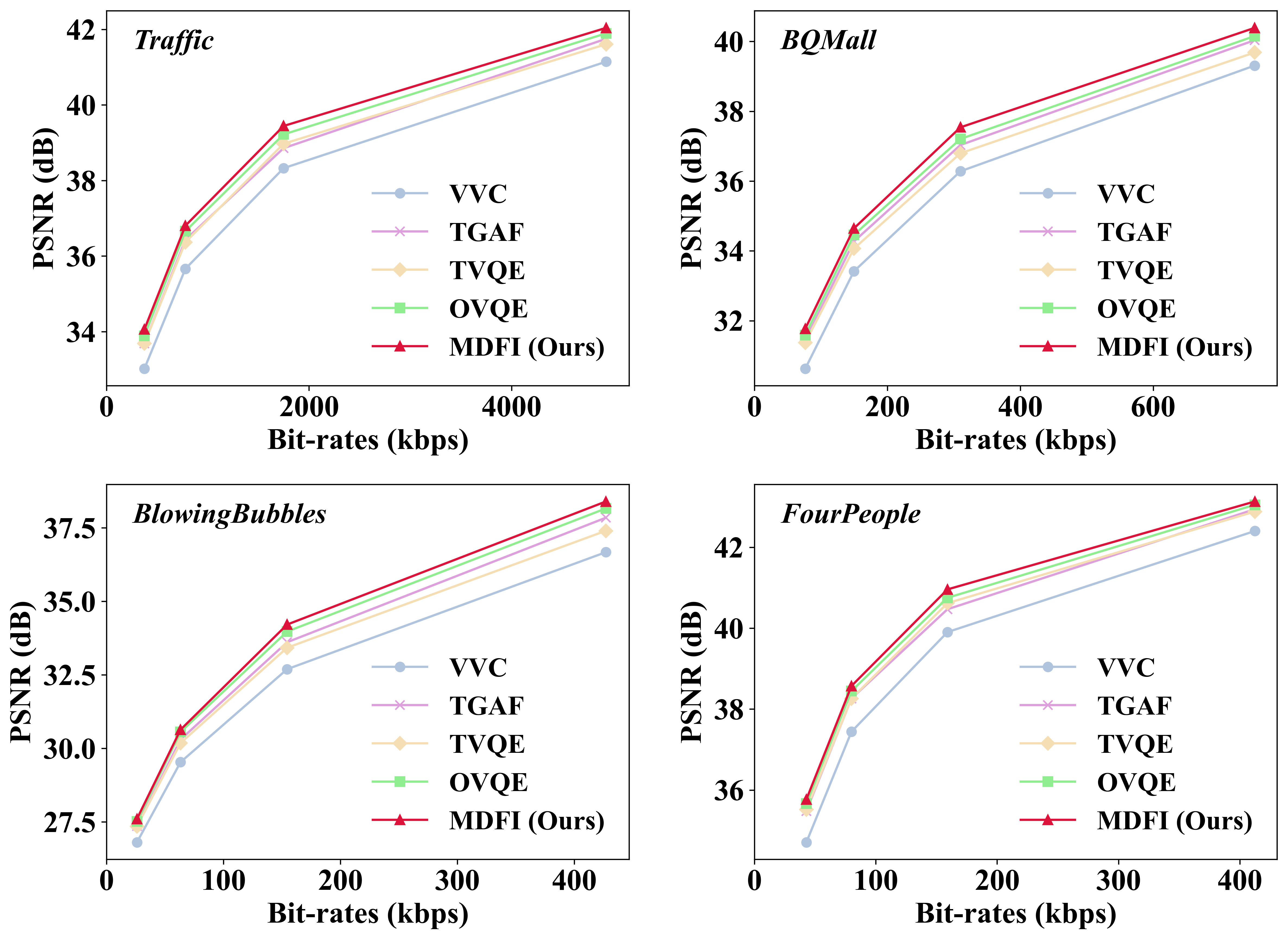}
    \caption{Rate-Distortion performance comparison on Traffic, BQSquare, RaceHorses, FourPeople.}
    \label{fig:rd_curve}
\end{figure}
\begin{figure}[!t]
    \centering
    \includegraphics[width=\linewidth]{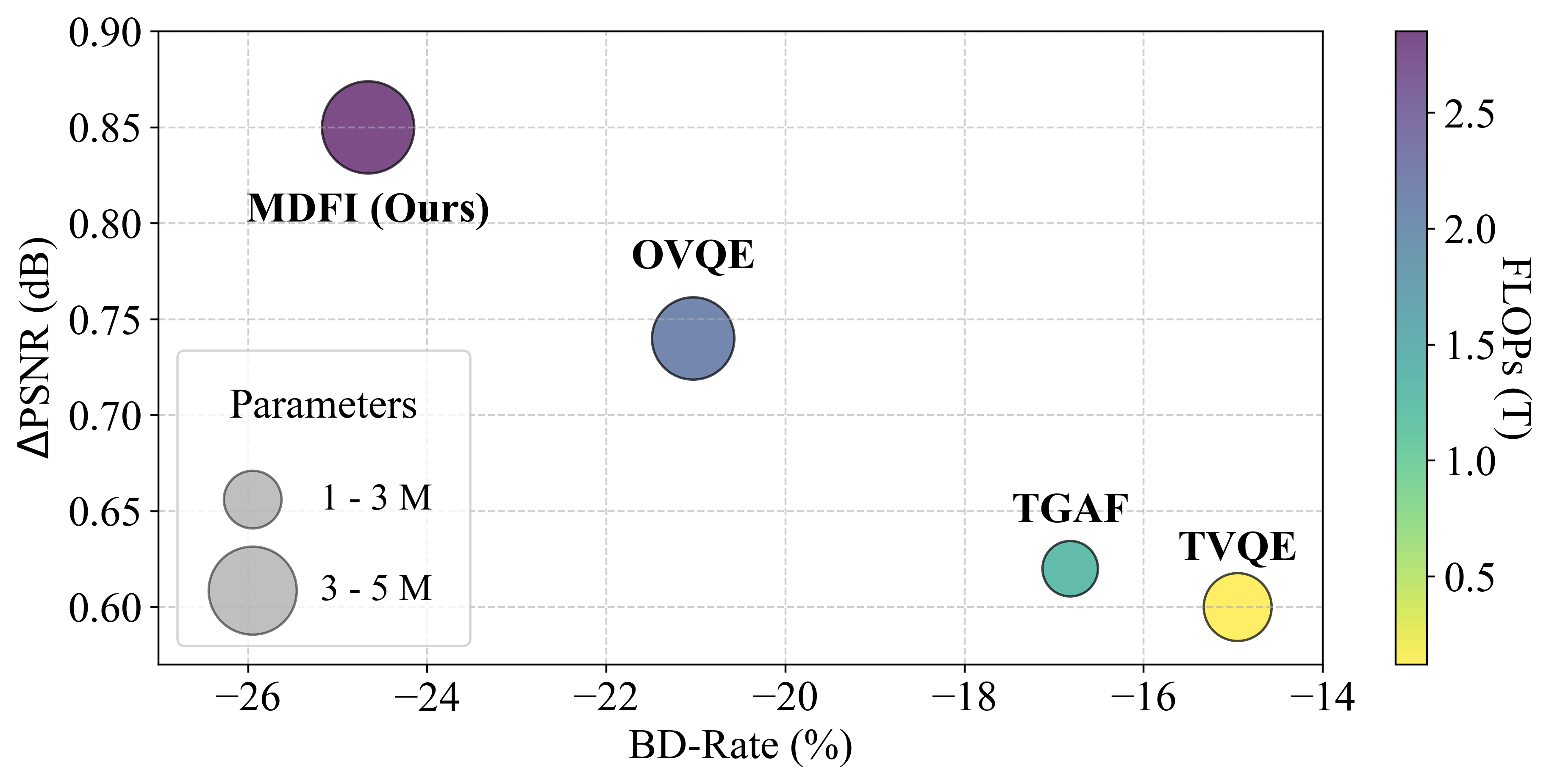}
    \caption{Trade-off analysis between quality improvement and model complexity.}
    \label{fig:complexity_visualization}
    \vspace{-0.2cm}
\end{figure}
\begin{table}[!t]
\scriptsize
\centering
\caption{Parameters, FLOPs, $\Delta$PSNR, and BD-rate comparisons of different methods at QP = 37.}
\label{tab:complexity}
\resizebox{\linewidth}{!}{%
\begin{threeparttable}
\begin{tabular}{lcccc} 
\hline
Model & Parameters (M) & FLOPs (T) & $\Delta$PSNR (dB) & BD-rate (\%) \\  \hline
TGAF~\cite{tgaf}   & 1.39          & 1.32      & 0.62              & -16.82   \\
TVQE~\cite{tvqe}           & 2.08          & 0.12      & 0.60              & -14.95   \\
OVQE~\cite{ovqe}        & 3.06          & 2.13      & 0.74              & -21.03   \\
{Wang et al~\cite{stff}} & {2.51} & {4.42} & {0.75} & {-21.35} \\
MDFI (Ours)                     & 3.84          & 2.85      & \textbf{0.85}     & \textbf{-24.66}  \\
\hline
\end{tabular}%
\end{threeparttable}
}
\end{table}


\textbf{Rate–Distortion Performance:} Table~\ref{tab:BDrate} compares the rate-distortion performance between competing methods in terms of BD-rate savings~\cite{bdrate}. As shown, MDFI achieves the best performance with a reduction of nearly 24.66\%, which is approximately 47\% higher than that of TVQE~\cite{tvqe} and TGAF~\cite{tgaf}. 
We also compare rate-distortion curves with other state-of-the-art methods on four test sequences. As can be seen from Fig.~\ref{fig:rd_curve}, for a similar bitrate, our method can obtain a higher PSNR, indicating our superior rate–distortion performance.

\textbf{Model Complexity:} Table~\ref{tab:complexity} and Fig.~\ref{fig:complexity_visualization} present an analysis of model complexity in terms of the number of parameters and FLOPs. Although the proposed method attains the best $\Delta$PSNR of 0.85 dB, it also requires the largest number of parameters and FLOPs, which negatively affects its inference speed. Despite this computational burden, MDFI remains particularly suitable for applications where performance is preferred over efficiency.
\begin{table}[!t]
\centering
\scriptsize
\caption{Ablation study under different \textit{out\_nc} and $N$ settings of MDFI, compared with baseline OVQE at QP=37.}
\label{tab:ablation_complexity}
\resizebox{\linewidth}{!}{%

\begin{tabular}{llccccc}
\hline
{Model} & Configuration & Parameters (M) & FLOPs (T) & {Time (ms)} & $\Delta$PSNR (dB) & {Ratio}\\ 
\hline
{MDFI} & \textit{out\_nc} = 64, $N = 16$ & 3.84 & 2.85 & {724} & \textbf{0.85}  & {3.35}\\
{MDFI\_S1} & \textit{out\_nc} = 64, $N = 1$ & 3.44 & 2.48 & {550} & 0.83 & {2.99}\\
{MDFI\_S2} & \textit{out\_nc} = 32, $N = 16$ & 2.22 & 1.61 & {501} & 0.74 & {2.17}\\
{MDFI\_S3} & \textit{out\_nc} = 32, $N = 1$ & \textbf{1.82} & \textbf{1.24} & {\textbf{326}} & 0.62 & {\textbf{2.00}}\\
{OVQE} & {Baseline} & {3.06} & {2.13} & {475} & {0.74} & {2.88}\\
\hline
\end{tabular}%
}
\begin{flushleft}
\tiny
\item \textit{out\_nc} denotes the number of channels in the model’s intermediate layers. Ratio = FLOPs / $\Delta$PSNR (lower is better).
\end{flushleft}
\vspace{-0.8cm}
\end{table}
\subsection{Ablation Study}

\textbf{Ablation Study on MDFI Configurations:} 
We analyze the effects of the number of SFTRBs ($N$) and channels (\textit{out\_nc}) on the performance–efficiency trade-off (Table~\ref{tab:ablation_complexity}). Increasing $N$ and \textit{out\_nc} improves $\Delta$PSNR but increases computational cost. To quantify this trade-off, we introduce FPRatio, defined as $\text{FLOPs}/\Delta\text{PSNR}$ (the lower, the better). While the full MDFI achieves the best performance, it has the highest Ratio, indicating lower efficiency. In contrast, lightweight variants (MDFI\_S2, MDFI\_S3) achieve better trade-offs with lower Ratio values. Notably, MDFI\_S2 attains comparable performance to OVQE with reduced complexity, whereas MDFI\_S3 provides the most efficient configuration with a slight performance drop. The full MDFI and MDFI\_S1 configurations maximize enhancement quality, making them ideal for cloud-based processing. Meanwhile, MDFI\_S2 matches baseline quality with 24.4\% fewer FLOPs (1.61T) to suit Smart TVs with AI accelerators, whereas MDFI\_S3 halves both computational cost and latency for moderate-budget edge devices.


\begin{table}[!t]
\centering
\scriptsize
\caption{Impact of different MDFI components on performance in terms of $\Delta$PSNR and $\Delta$SSIM at QP=37.}
\label{tab:ablation_component}

\begin{tabular}{l c c c c c}
\hline
FPFT & STFF & GMFF & FQE & $\Delta$PSNR & $\Delta$SSIM \\ 
\hline
           &            &            & \checkmark & 0.35 & 0.74 \\
           &            & \checkmark & \checkmark & 0.46 & 1.00 \\
           & \checkmark & \checkmark & \checkmark & 0.74 & 1.49 \\
\checkmark & \checkmark & \checkmark & \checkmark & \textbf{0.85} & \textbf{1.68} \\
\hline
\end{tabular}
\vspace{-0.5cm}
\end{table}

{\textbf{Ablation Study on Different Components in MDFI:}
To evaluate the contribution of each component in the MDFI architecture, we analyze the restoration performance in terms of $\Delta$PSNR and $\Delta$SSIM under different combinations of FPFT, STFF, GMFF, and FQE. 

As shown in Table~\ref{tab:ablation_component}, the complete MDFI configuration, with all modules integrated, yields the best results, improving $\Delta$PSNR by 0.85 dB and $\Delta$SSIM by 1.68 $\times$ $10^{-2}$. It outperforms the second-best variant (without FPFT) by 0.11 dB and 0.19, confirming the FPFT module's effectiveness. The FQE-only design still achieves a 0.35 dB gain in PSNR, highlighting the potential of a lightweight CVQE approach.
}

\begin{table}[!t]
\centering
\scriptsize
\caption{Overall evaluation for \(\Delta\)PSNR (dB), \(\Delta\)SSIM  ($\times 10^{-2}$) , \(\Delta\)VMAF and \(\Delta\)LPIPS ($\times 10^{-2}$)  over standard test sequences at QP=37.}
\label{tab:ablation_metrics}

\begin{tabular}{lcccc}
\hline
Model      & \(\Delta\)PSNR $\uparrow$ & \(\Delta\)SSIM $\uparrow$ &  \(\Delta\)VMAF $\uparrow$ & \(\Delta\)LPIPS $\downarrow$\\ \hline
TGAF [11]       & 0.62    & 1.23    & 2.33     &  -0.63     \\
TVQE [12]      & 0.60    & 1.19    & 2.86     &  -0.49     \\
OVQE [13]      & 0.74    & 1.49    & 2.96     &  -1.01 \\
Wang et al [14]       & 0.75	 & 1.48    & 3.13     &  -0.79 \\
MDFI (Our) & \textbf{0.85}   & \textbf{1.68}    & \textbf{3.79}     &    \textbf{-1.17} \\ \hline  
\end{tabular}
\vspace{-0.6cm}
\end{table}
{\textbf{Perceptual Quality Analysis:} To comprehensively evaluate visual quality, we incorporate perceptual metrics (VMAF, LPIPS) in Table~\ref{tab:ablation_metrics}. As shown, MDFI consistently achieves the highest $\Delta$VMAF and lowest $\Delta$LPIPS, confirming its ability to enhance perceptual quality and suppress artifacts.
Compared with existing methods, gains in perceptual metrics are more pronounced than distortion-based ones, showing MDFI’s strength in restoring visually pleasing details under high compression while balancing distortion reduction and perceptual enhancement.
}

\section{Conclusions}
\label{sec:conclusions}
In this paper, we propose a novel multi-domain feature integration model that not only captures spatiotemporal correlations between adjacent reconstructed frames but also leverages prediction information directly from the compressed domain for H.266/VVC, the latest video coding standard, to enhance compressed video quality. By introducing the effective FPFT module, our approach integrates multi-domain features and exploits long-term dependencies across multiple frames and prediction signals. Furthermore, we introduce a dedicated dataset that includes not only pairs of raw–reconstructed sequences but also prediction frames extracted from the H.266/VVC bitstream, providing a valuable foundation for future compressed video quality enhancement research. Extensive experiments demonstrate that our method significantly improves video quality and reduces compression artifacts, outperforming existing state-of-the-art methods in both objective and subjective evaluations.


\bibliographystyle{IEEEtran}
\bibliography{paper.bib}
\end{document}